\documentclass[aoas,MSNbibl,nameyear,rotating,dvips]{arximspdf}
\usepackage{multirow}
\usepackage{graphicx}

\doi{10.1214/11-AOAS488}
\volume{5}
\issue{4}
\pubyear{2011}
\firstpage{2359}
\lastpage{2384}

\newcommand{\eqref}[1]{(\ref{#1})}

\begin{document}
\begin{frontmatter}

\title{Modeling of the HIV infection epidemic in the Netherlands: A
multi-parameter evidence synthesis~approach\thanksref{TT1}}
\runtitle{MPES modeling of HIV prevalence in the Netherlands}

\begin{aug}
\author[A]{\fnms{Stefano} \snm{Conti}\corref{}\ead[label=e1]{stefano.conti@hpa.org.uk}},
\author[B]{\fnms{Anne M.} \snm{Presanis}},
\author[C]{\fnms{Maaike G.} \snm{van Veen}},
\author[C]{\fnms{Maria} \snm{Xiridou}},
\author[D]{\fnms{Martin C.} \snm{Donoghoe}},
\author[D]{\fnms{Annemarie} \snm{Rinder Stengaard}}
\and
\author[E]{\fnms{Daniela} \snm{De Angelis}}
\runauthor{S.~Conti et al.}
\affiliation{Health Protection Agency,
Medical Research Council Biostatistics Unit,
National Institute for Public Health and the Environment,
National Institute for Public Health and the Environment,
World Health Organization Regional Office for Europe,
World Health Organization Regional Office for Europe, and
Health Protection Agency and Medical Research Council Biostatistics Unit}
\address[A]{S.~Conti\\
Statistics Unit\\Centre for Infections\\Health
Protection Agency\\61 Colindale Avenue\\London NW9 5EQ\\
United Kingdom\\
\printead{e1}} 
\address[B]{A. M. Presanis\\
MRC Biostatistics Unit\\Institute of Public
Health\hspace*{6pt}\\University Forvie Site\\Robinson Way\\Cambridge CB2
0SR\\
United Kingdom}
\address[C]{M. G. van Veen\\
M. Xiridou\\
National Institute for Public Health\\ \quad and the
Environment\\PO Box 1\\3720 BA Bilthoven\\The Netherlands}
\address[D]{M. C. Donoghoe\\
A. Rinder Stengaard\\
World Health Organization\\Regional Office for
Europe\\Scherfigsvej 8\\DK-2100 Copenhagen \O\\
Denmark}
\address[E]{D. De Angelis\\
Statistics Unit\\Centre for Infections\\Health
Protection Agency\\61 Colindale Avenue\\London NW9 5EQ\\
United Kingdom\\
and\\
MRC Biostatistics Unit\\Institute of Public
Health\\University Forvie Site\\Robinson Way\\Cambridge CB2
0SR\\
United Kingdom}
\end{aug}
\thankstext{TT1}{Supported by the World Health Organization Regional
Office for Europe (WHO/Europe) and  Medical Research Council
Grants U.1052.00.007 and G0600675.}

\received{\smonth{6} \syear{2010}}
\revised{\smonth{5} \syear{2011}}

%
\vspace*{3pt}
\begin{abstract}
Multi-parameter evidence synthesis (MPES) is receiving growing
attention from the epidemiological community as a coherent and
flexible analytical framework to accommodate a disparate body of
evidence available to inform disease incidence and prevalence
estimation. MPES is the statistical methodology adopted by the
Health Protection Agency in the UK for its annual national
assessment of the HIV epidemic, and is acknowledged by the World
Health Organization and UNAIDS as a valuable technique for the
estimation of adult HIV prevalence from surveillance data. This
paper describes the results of utilizing a Bayesian MPES approach to
model HIV prevalence in the Netherlands at the end of 2007, using an
array of field data from different study designs on various
population risk subgroups and with a varying degree of regional
coverage. Auxiliary data and expert opinion were additionally
incorporated to resolve issues arising from biased, insufficient or
inconsistent evidence. This case study offers a demonstration of the
ability of MPES to naturally integrate and critically reconcile
disparate and heterogeneous sources of evidence, while producing
reliable estimates of HIV prevalence used to support public health
decision-making.
\end{abstract}

%

\end{frontmatter}

\section{Introduction}
\label{secintro}
Refining and advancing the current understanding of the dynamics of
the HIV epidemic attracts a continued interest from the
epidemiological and medical community. Both national and international
public health institutes recognize the importance of improving current
methods to monitor HIV prevalence, as this constitutes a key input to
inform public health-care policies and resource allocation.

A number of approaches have been proposed in the statistical
literature, starting from the back-calculation method
[\citet{brkgai88}], initially devised to obtain an estimate of HIV
prevalence. The most popular estimation methods (so-called ``direct'')
typically rely on evidence specifically around HIV prevalence
[\citet{gscjhn94};
\citet{ptrncl97};
\citet{hwlhst98};
\citet{krnkhr98};
\citet{rmnalv02};
\citet{mcgclf06}]. In
broad terms, direct methods assume a target population of size
$N=\sum_gN_g$ to be divided into mutually exclusive subgroups
$g=1,\ldots,G$ of corresponding size $N_g$. Each subgroup is
characterized by a given degree of risk behavior and consists of
$N_g(1-\pi_g)$ uninfected and $N_g\pi_g$ infected individuals, where
$\pi_g$ denotes the unknown subgroup-specific HIV
prevalence. Prevalent cases $N_g\pi_g$ can in turn be split into
$N_g\pi_g\delta_g$ diagnosed and $N_g\pi_g(1-\delta_g)$ undiagnosed
individuals, as determined by the (unknown) proportion $\delta_g$ of
HIV positive cases diagnosed within each subgroup. Provided enough
cross-sectional surveillance- or survey-based information is available
to estimate subgroup sizes and parameters, the number of
subgroup-specific diagnosed and undiagnosed prevalent cases can be
inferred by multiplying corresponding estimates of $N_g$ and~$\pi_g$
with $\delta_g$ and $1-\delta_g$, respectively. These in turn can be
summed across subgroups to obtain a point estimate of the total number
of HIV infections in the population.

While at a first glance appealing, direct methods suffer from both
conceptual and practical complications. Data may: (i) be insufficient
to inform directly relevant parameters, like prevalence in
hard-to-reach subgroups; (ii) relate to individuals matching multiple
risk profiles; and/or (iii) be affected by selection and reporting
biases. Without the inclusion of supplementary evidence, these
problems are normally tackled via unverifiable assumptions, ad-hoc
adjustments and/or removal of selected data
[\citet{gouad08}]. Moreover, the common practice of using only as many
items of (highest quality) evidence as the number of parameters of
interest is hardly justified under a decision-making
perspective. Decisions around research prioritization and service
provision are more rationally and robustly taken when driven by
a~comprehensive, rather than selective, use of available information
[\citet{clxscl02}], provided the varying degree of accuracy of the
components of the evidence base is correctly recognized and taken into
account in the analysis.

Conversely, \textit{multi-parameter evidence synthesis} (henceforth
MPES) offers a coherent analytical framework designed to make rational
and exhaustive use of the whole body of information available
[\citet{adsu06}], thus circumventing the above shortcomings. A
disparate pool of evidence is naturally accommodated within a MPES
model structure through its formal specification of the relationships
between data and parameters, which dictate how (\textit{direct})
evidence on the parameters of interest can be supplemented by
(\textit{indirect}) information available on arbitrarily complex
functions of those parameters. A MPES approach thus presents a number
of advantages over direct methods: first, since it incorporates more
data, a MPES model is expected to produce more accurate parameter
estimates. Consequently, the inferences it produces correctly reflect
the uncertainty surrounding the whole evidence base. Moreover, where
there are more data points than estimands, MPES flags any
inconsistency potentially affecting a collection of heterogeneous
items of data. These conflicts are important to detect, as they may
highlight biases in, or misinterpretations of, the data, which can be
then addressed.

As an analytical perspective, MPES has in recent years rapidly gained
a~foot in medical statistics, health technology assessment and
epidemiological modeling of infectious diseases like HIV and hepatitis
C [\citet{wltads05};
\citet{gouad08};
\citet{presdean08};
\citet{swdan08};
\citet{deanswe09}]. Since
2005 the UK Health Protection Agency employs a MPES approach to
estimate diagnosed and undiagnosed HIV prevalences in the UK using
data from routine surveillance and ad-hoc surveys
[HIV \& STI Department (\citeyear{HPA05,HPA06,HPA07,HPA08,HPA09})]. These evidence synthesis
exercises have typically been carried out from a Bayesian perspective,
due to its computational convenience, coherent decision-theoretic
foundation and automatic synthesis between empirical and
prior/subjective information.

This paper describes the development of a Bayesian MPES model to
estimate HIV prevalence in different population subgroups and areas
across the Netherlands, through reliance on its national surveillance
network and an array of regional registries and surveys. The proposed
model produces estimates of prevalence, proportions diagnosed and
sizes for a number of pre-defined subgroup profiles at risk of HIV
infection within the target population of 15- to 70-year old
individuals living in the Netherlands in 2007. The paper is organized
as follows: Section \ref{seces} formally defines the MPES approach
adopted. Section \ref{secdata} describes the body of evidence
compiled by the National Institute for Public Health and the
Environment in the Netherlands to enable estimation. Section
\ref{secmod} details the MPES model building process, and results are
illustrated in Section \ref{secres}. Model criticism and concluding
remarks are outlined in Section \ref{secdisc}.
\section{The synthesis of evidence}
\label{seces}

The practice of synthesising evidence from multiple sources, through
the combination of direct and/or indirect information from differently
designed studies, dates well before dedicated work emerged under an
explicit MPES header. Besides the vast body of literature on
meta-analytis [see \citet{sutabr00}], of which MPES represents an
extension, a methodological stepping stone in the subject of collating
direct and indirect evidence is widely recognized to be the Confidence
Profile Method [\citet{edhas92}]. Instances of complex synthesis
include, but are not limited to, indirect and mixed treatment
comparisons [e.g., \citet{dompar99};
\citet{sonalt03};
\citet{luad04};
\citet{calad05}],
cross-design synthesis [\citet{drosil93};
\citet{benhar00}], hierarchical
models [extensively reviewed in, e.g., \citet{sutabr00};
\citet{adcli02};
\citet{whi02};
\citet{gelhil07}], Bayesian melding
[\citet{pooraf00};
\citet{fueraf05};
\citet{alkraf07}], bias adjustment
[\citet{spiegbes03};
\citet{wolmen04};
\citet{turspieg09}] and multiple/surrogate
endpoint synthesis [\citet{berhoa98};
\citet{nammen03};
\citet{burmol04}]. These
examples attempt to integrate separate sources of evidence to draw
inferences that are not only more efficient than those instead
obtained from a selective ``best data'' approach, but also consistent
with \textit{all} available information.

Formally, a MPES setup follows closely the characterization of the
Confidence Profile Method: assume interest lies in learning about $I$
\textit{basic} parameters
$\bolds\vartheta=(\vartheta_1,\ldots,\vartheta_I)$, and that for
estimation purposes $n$ data points (i.e., sufficient statistics)
$\mathbf y=(y_1,\ldots,y_n)$ have been separately collected. Any
data point may inform either a basic parameter $\vartheta_i$ or some
\textit{functional} parameter $\psi_j=\psi_j(\bolds\vartheta),\
j=1,\ldots,J$, which can be expressed as a function of known form of
the basic parameters. Data unbiasedly reporting on basic parameters
are normally referred to as ``direct'' evidence; samples informing
functional parameters are also included in the evidence base, in that
they provide ``indirect'' evidence about their defining basic
parameters. Indicating with $L_r(\bolds\vartheta;y_r)$ the
likelihood contribution from $y_r$ to (elements of) the basic
parameter vector $\bolds\vartheta$, from the independence of
elements in~$\mathbf y$ the full likelihood model
%
\begin{equation}
\label{eqL}
L(\bolds\vartheta;\mathbf y)=\prod_{r=1}^nL_r(\bolds\vartheta;y_r)
\end{equation}
follows.

Within a classical framework, specification of \eqref{eqL} is
sufficient to obtain, typically via maximum likelihood, estimates
$\hat{\bolds\vartheta}$ of the basic parameters and therefore of
the $J$ functional parameters
$\hat{\bolds\psi}_j=\psi_j(\hat{\bolds\vartheta})$. Additionally,
under a~Bayesian perspective, prior (imperfect or even scarce)
knowledge around the basic parameters, as expressed through some joint
prior distribution~$p(\bolds\vartheta)$, may be updated in the
light of the observed data into a posterior distribution~$p(\bolds\vartheta|\mathbf y)$ summarizing all information
around $\bolds\vartheta$ (and thus $\bolds\psi$): that is,
\[
p(\bolds\vartheta|\mathbf y)\propto
p(\bolds\vartheta)L(\bolds\vartheta;\mathbf y).
\]
As in
recent MPES modeling work, a Bayesian approach is here proposed since
its prior-to-posterior updating mechanism naturally corresponds to the
spirit, typical of MPES, of synthesizing multiple items of
evidence. Furthermore, the resulting posterior distribution fully
reflects both the sampling variability affecting such evidence and the
parameter uncertainty surrounding the model.

\section{The HIV surveillance network in the Netherlands}
\label{secdata}

$\!\!$In line with \citet{gouad08} and \citet{presdean08}, and compatibly
with the socio-demographic coverage and resolution of available data,
the population living in the Netherlands at the end of 2007 was
classified by mutually exclusive subgroups and areas of
residence. Subgroups are defined as follows:

\begin{longlist}[(5)]
\item[(1)] men who have sex with men ($\mathrm{MSM}$), who have ($\mathrm{MSM}_{\mathrm{STI}}$) or
have not ($\mathrm{MSM}_{\overline{\mathrm{STI}}}$) attended a sexually-transmitted
infections (STI) clinic in 2007;
\item[(2)] intravenous drug users (IDU);
\item[(3)] female sex workers (FSW);
\item[(4)] heterosexuals attending an STI clinic ($\mathrm{STI}$), further divided
into Sub-Sa\-haran Africans ($\mathrm{SSA}_{\mathrm{STI}}$), Caribbeans ($\mathrm{CRB}_{\mathrm{STI}}$)
and nonmigrants ($\mathrm{WST}_{\mathrm{STI}}$);
\item[(5)] heterosexuals not attending an STI clinic (thus supposedly at low
risk of infection), further divided into Sub-Saharan Africans
($\mathrm{SSA}_{\overline{\mathrm{STI}}}$), Caribbeans ($\mathrm{CRB}_{\overline{\mathrm{STI}}}$) and
nonmigrants ($\mathrm{WST}_{\overline{\mathrm{STI}}}$).
\end{longlist}

Let $\mathcal G$ denote the set collecting the above
subgroups. Broader groups may be defined by merging selected risk
categories in $\mathcal G$, such as migrants from HIV-endemic areas
($\mathrm{MGR}\doteq \mathrm{MGR}_{\mathrm{STI}} \cup \mathrm{MGR}_{\overline{\mathrm{STI}}}$) either attending
($\mathrm{MGR}_{\mathrm{STI}}\doteq \mathrm{SSA}_{\mathrm{STI}} \cup \mathrm{CRB}_{\mathrm{STI}}$) or not attending
($\mathrm{MGR}_{\overline{\mathrm{STI}}}\doteq \mathrm{SSA}_{\overline{\mathrm{STI}}}\cup
\mathrm{CRB}_{\overline{\mathrm{STI}}}$) an STI clinic; likewise, nonmigrant population
clusters ($\mathrm{WST}\doteq \mathrm{WST}_{\mathrm{STI}} \cup \mathrm{WST}_{\overline{\mathrm{STI}}}$) may be
similarly defined. Here it is assumed that subgroups in $\mathcal G$
are ranked by decreasing risk of infection, so that individuals
matching multiple risk profiles are allocated into the one highest
ranked: for instance, FSW who are at the same time IDU would be
classified as IDU.

Group and gender specific estimates of key parameters are derived for
three geographic regions: Amsterdam ({A}), Rotterdam
({R}) and the rest of the country ({O}). Let $N_r$
indicate the total population residing in region~$r$, assumed known
from census statistics, and $N_{r,g}=\rho_{r,g}N_r$ the unknown (to be
estimated) \textit{absolute} size of subgroup $g\in\mathcal G$
therein. Basic parameters of interest consist of \textit{relative}
subgroup size $\rho_{r,g}$, HIV prevalence $\pi_{r,g}$ and proportion
diagnosed with HIV $\delta_{r,g}$ for each combination of 9 subgroups
$g$ and 3 regions $r$. With group-specific estimates being sought by
gender (and by STI clinic attendance status for MSM) except for the
female-only FSW, the total number of independent basic estimands thus
amounts to
\[
3\times(\overbrace{9\times2-1-2}^{\#\{\rho_{r,g}\}}+\overbrace{9\times
2-1}^{\#\{\pi_{r,g}\}}+\overbrace{9\times2-1}^{\#\{\delta_{r,g}\}})=147,
\]
given that regional subgroup proportions add up to 1 for each gender:
$\sum_g\rho_{r,g}=1\ \forall r$.

The HIV surveillance network in place in the Netherlands provides
sufficient information to infer basic parameters for most
region-subgroup combinations. However, data are partly lacking on
proportions diagnosed (notably among migrant subgroups) and more
generally outside main urban areas. This lack of information
complicates, and in certain cases prevents, estimation of relevant
basic parameters, so that a direct approach in the spirit of that
outlined in Section \ref{secintro} would be inapplicable. On the
other hand, an array of registry-based and ad-hoc surveys effectively
targeting functional parameters is available to supplement, from a
MPES perspective, the available direct data, therefore compensating
for the poor evidence on some basic parameters. The overall data set
consists of 194 items of data: 65 from Amsterdam, 60 from Rotterdam
and 69 from the rest of the Netherlands.

\begin{figure}

\includegraphics{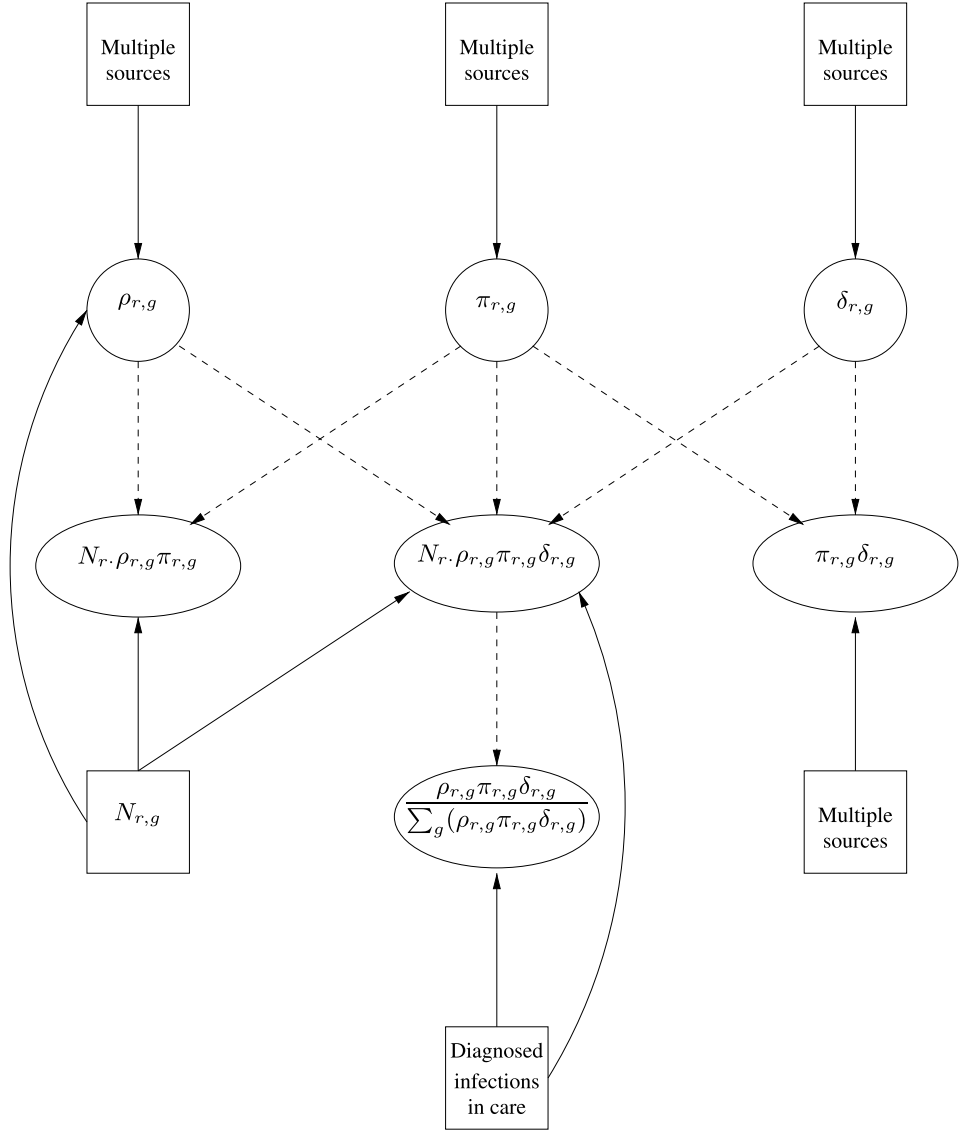}

\caption{Schematic representation of the evidence network informing
epidemiological parameters in the MPES HIV model: different
samples (squares) provide data-based information (solid arrows)
around key basic (circles) and functional (ellipses) estimands,
where the latter are functionally related (dashed arrows) to the
former (i.e., $\rho_{r,g}, \pi_{r,g}\mbox{ and }\delta_{r,g}$).}
\label{figinf}
\end{figure}

Table \ref{taba} details the data collected to directly or indirectly
inform HIV epidemic descriptors in the Amsterdam area; the network of
surveillance and survey data capturing the HIV epidemic in Rotterdam
and the rest of the Netherlands is reported as 
\ref{supdataro}. The full array of data shows the extent of coverage of
national surveillance and highlights the links between basic and
functional parameters. Figure \ref{figinf} sketches the flow of
information within the network of evidence, which is described below.

%
\begin{sidewaystable}
\tabcolsep=0pt
\tablewidth=\textwidth
\caption{Evidence supporting HIV prevalence estimation in
Amsterdam ($N_m=284\mbox{,}002$, $N_f=284\mbox{,}067$); letters in
brackets link to data sources as detailed in Section
\protect\ref{secdata}}
\label{taba}
\begin{tabular*}{\textwidth}{@{\extracolsep{\fill}}lcccccc@{}}
\hline
 & &\multicolumn{3}{c}{\textbf{Basic parameters}} & \multicolumn{2}{c@{}}{\textbf{Functional parameters}}\\[-5pt]
 & &\multicolumn{3}{c}{\hrulefill} & \multicolumn{2}{c@{}}{\hrulefill}\\
\textbf{Group}&\textbf{Subgroup} & $\bolds\rho$ & $\bolds\pi$ & $\bolds\delta$ & $\bolds{\pi\delta}$ & $\bolds\mu$\tabnoteref{tbl1a}\\
\hline
\multirow{2}{*}{{MSM}} & {{STI}} & $2\mbox{,}495/N_m=0.009$ (f) &
606/2\mbox{,}723${}={}$0.223\tabnoteref{fn:ams6}
(f) & 79/85${}={}$0.929 (g) & &\\
& {All} & 73/776${}={}$0.094 (a) & & 48/547${}={}$0.088 (s) & &
2\mbox{,}827\tabnoteref{tbl1c} (h)\\[3pt]
\multirow{2}{*}{{IDU}} & {M} &
(720--1\mbox{,}120)$/N_m=0.003\mbox{--}0.004$ (i) & 45/167${}={}$0.269\tabnoteref{fn:ams3} (k) &
31/45${}={}$0.689\tabnoteref{fn:ams4} (k) &
37/196${}={}$0.189 (p) & 99\tabnoteref{fn:ams5}
(h)\\
& {F} & (180--280)$/N_f=6.34E\mbox{--}04\mbox{--}9.85E\mbox{--}04$ (i) &
6/30${}={}$0.200\tabnoteref{fn:ams3} (k) & 3/6${}={}$0.500\tabnoteref{fn:ams4} (k) &
20/88${}={}$0.227 (p) & 64\tabnoteref{fn:ams5} (h)\\[3pt]
{FSW} & {F} & $7\mbox{,}440/N_f=0.026$ (q) & 3/148${}={}$0.020 (q) &
0/3${}={}$0 (q) & &\\[3pt]
\multirow{2}{*}{$\mathrm{WST}_{\mathrm{STI}}$} & {M} & $5\mbox{,}702/N_m=0.020$ (f) &
10/5\mbox{,}526${}={}$0.002\tabnoteref{fn:ams6} (f) & $10/(N_{g,m}\pi_{g,m})$ (f) & &\\
& {F} & $6\mbox{,}586/N_f=0.023$ (f) & 7/6\mbox{,}402${}={}$0.001\tabnoteref{fn:ams6}
(f) & $7/(N_{g,f}\pi_{g,f})$ (f) & &\\[3pt]
\multirow{2}{*}{$\mathrm{SSA}_{\mathrm{STI}}$} & {M} & $261/N_m=0.001$ (f) &
7/237${}={}$0.030\tabnoteref{fn:ams6} (f) & $7/(N_{g,m}\pi_{g,m})$ (f) & &\\
& {F} & $158/N_f=0.001$ (f) & 10/151${}={}$0.066\tabnoteref{fn:ams6} (f)
& $10/(N_{g,m}\pi_{g,f})$ (f) & &\\ [3pt]
\multirow{2}{*}{$\mathrm{CRB}_{\mathrm{STI}}$} & {M} & $899/N_m=0.003$ (f) &
4/855${}={}$0.005\tabnoteref[b]{fn:ams6} (f) & $4/(N_{g,m}\pi_{g,m})$ (f) & &\\
& {F} & $771/N_f=0.003$ (f) & 6/753${}={}$0.008\tabnoteref[b]{fn:ams6} (f)
& $6/(N_{g,f}\pi_{g,f})$ (f) & &\\
[3pt]
\multirow{2}{*}{$\mathrm{SSA}_{\overline{\mathrm{STI}}}$} & {M} &
$9\mbox{,}434/N_m=0.033$\tabnoteref[g]{fn:ams7} (o) & 1/129${}={}$0.008 (l) &
& & 173\tabnoteref[h]{tbl1h} (h)\\
& {F} & $8\mbox{,}233/N_f=0.029$\tabnoteref[g]{fn:ams7} (o) & & 0/50${}={}$0 (l) &
& $252^{\mathrm{h}}$ (h)\\[3pt]
\multirow{2}{*}{$\mathrm{CRB}_{\overline{\mathrm{STI}}}$} & {M} &
$31\mbox{,}200/N_m=0.110$\tabnoteref[g]{fn:ams7} (o) & 1/215${}={}$0.005 (l) & & &
$137^{\mathrm{h}}$ (h)\\
& {F} & $36\mbox{,}468/N_f=0.128$\tabnoteref[g]{fn:ams7} (o) & 1/252${}={}$0.004
(l) & & & $111^{\mathrm{h}}$ (h)\\[3pt]
\multirow{2}{*}{{{STI}}} & {M} & & 8/2\mbox{,}135${}={}$0.004 (g) &
$2/8=0.250^{\mathrm{e}}$ (g) & &\\
& {F} & & 7/2\mbox{,}580${}={}$0.003 (g) & $3/7=0.429^{\mathrm{e}}$ (g) & &\\
\hline
\end{tabular*}
\end{sidewaystable}
\setcounter{table}{0}
\begin{sidewaystable}
\tabcolsep=0pt
\tablewidth=\textwidth
\caption{(Continued)}
\begin{tabular*}{\textwidth}{@{\extracolsep{\fill}}lcccccc@{}}
\hline
 & &\multicolumn{3}{c}{\textbf{Basic parameters}} & \multicolumn{2}{c@{}}{\textbf{Functional parameters}}\\[-5pt]
 & &\multicolumn{3}{c}{\hrulefill} & \multicolumn{2}{c@{}}{\hrulefill}\\
\textbf{Group}&\textbf{Subgroup} & $\bolds\rho$ & $\bolds\pi$ & $\bolds\delta$ & $\bolds{\pi\delta}$ & $\bolds\mu$\tabnoteref[a]{tbl1a}\\
\hline
\multirow{2}{*}{{Mixed}} & {M} & & & & & 145\tabnoteref[i]{fn:ams9} (h)\\
& {F} & & & & & 207\tabnoteref[i]{fn:ams9} (h)\\[3pt]
{Pregnant} & {Nonmigrant} & & 4/13\mbox{,}097${}={}$1E--04 (m) &
3/4${}={}$0.750 (m) & &\\
{women} & {Migrant} & & 27/3\mbox{,}413${}={}$0.008 (m) & 21/27${}={}$0.778
(m) & &\\
\hline
\end{tabular*}
\tabnotetext[a]{tbl1a}{{Total Amsterdam residents for each gender are estimated (source: CBS)
at 3\mbox{,}522 (M) and 660 (F), also including 141 and 26 cases of unknown
exposure respectively (source: SHM).}}
\tabnotetext[b]{fn:ams6}{{Data inform minimum prevalences, due to
278/2\mbox{,}495${}={}$0.111 (STI MSM), 182/5\mbox{,}702${}={}$0.032 (M WST), 186/6\mbox{,}586${}={}$0.028 (F
WST), 26/261${}={}$0.100 (M SSA), 9/158${}={}$0.057 (F SSA), 45/899${}={}$0.050 (M CRB)
and 19/771${}={}$0.025 (F CRB) STI users opt-out fractions.}}
\tabnotetext[c]{tbl1c}{{Registered cases are underestimated by a
91/98${}={}$0.929 fraction (source: Schorer Monitor).}}
\tabnotetext[d]{fn:ams3}{{Data
inform maximum prevalence.}}
\tabnotetext[e]{fn:ams4}{{Data inform minimum proportions diagnosed.}}
\tabnotetext[f]{fn:ams5}{{Registered cases are
underestimated by uninformed gender-specific fractions.}}
\tabnotetext[g]{fn:ams7}{{Counts also include STI clinic users,
but exclude illegal immigrants.}}
\tabnotetext[h]{tbl1h}{{Recorded infections include both STI clinic
attending and nonattending immigrants.}}
\tabnotetext[i]{fn:ams9}{{Mixture of respectively male and female registered infections
for IDU, FSW (F only), nonmigrant STI clinic users and other.}}
\end{sidewaystable}

%
%

\subsection{Relative subgroup sizes}
\label{ssecrho}

Official figures from Statistics Netherlands (CBS, source o) in Table
\ref{taba} provide absolute frequencies $N_r$ of regional
population sizes. Subgroup sizes for migrants not attending an STI
clinic, additionally obtained from CBS, are used to estimate
proportions $\rho_{r,g}$ for $g\in \mathrm{MGR}_{\overline{\mathrm{STI}}}$. It should be
noted, however, that CBS statistics neither track illegal entries into
the country nor distinguish between migrants attending an STI
clinic. This implies the following: (i) a downward bias undermining, to an
undocumented extent, migrant-related figures; and (ii) the need to
decouple STI clinic users from nonusers. Details of how these biases
are accommodated within the MPES modeling framework are outlined in
Section~\ref{secmod}.

Broad MSM subgroup sizes are derived from regional published
population studies (Amsterdam and Rotterdam Health Monitors, sources
a
and r) and random population samples\vadjust{\goodbreak} (RNG, source s; Pienter Project,
source~t). These data, however, enable unbiased estimation only of
$\rho_{\mathrm{A},\mathrm{MSM}}$, as they either under- or over-report absolute (and
hence relative) sizes of MSM subgroups living outside Amsterdam. These
data are then used to inform minimum and maximum values of the
underlying subgroup sizes. The municipal registry of opiate and
methadone users (source i) presents the same problem when used to
inform IDU prevalence in Amsterdam; the size of the IDU population
elsewhere is estimated unbiasedly through the Pienter study and a
municipal report on addiction and homelessness in Rotterdam (source
t).

Direct estimates of subgroup sizes from all STI clinic
attending-subgroups are obtained from the national registry of STI
clinic consultations (SOAP, source f). Finally, unlinked anonymous (UA)
HIV surveys (source q), reporting results from HIV antibody testing of
saliva samples from FSW in Amsterdam and Rotterdam, are available to
directly inform $\rho_{r,\mathrm{FSW}}$ for $r=\mathrm{A},\mathrm{R}$. A published study on FSW
in The Hague (source u) is utilized to inform a~range for the
frequency of FSW in the rest of the country.
\subsection{HIV prevalences}
\label{ssecpi}

Evidence around HIV prevalence is more fragmented than that on
subgroup sizes. Information relating to MSM individuals is sparse,
with the only direct source of evidence on $\pi_{\mathrm{A}, \mathrm{MSM}}$ consisting
of the Amsterdam Cohort Study (source s). Information outside urban
concentrations is indirectly derived through data on
\textit{diagnosed} prevalence (i.e., $\pi_{\mathrm{O}, \mathrm{MSM}}\delta_{\mathrm{O}, \mathrm{MSM}}$)
from the Schorer Monitor (source e), that is, the national institute
responsible for the coordination of primary HIV/STI prevention
policies targeting MSM in the Netherlands, and the Pienter
Project. Since these sources are biased downward and upward,
respectively, they provide upper and lower limits for the product
$\pi_{\mathrm{O},\mathrm{MSM}}\delta_{\mathrm{O},\mathrm{MSM}}$. Moreover, the Pienter data set also
supplies information on diagnosed low-risk prevalence outside main
urban areas
($\pi_{\mathrm{O},\mathrm{WST}_{\overline{\mathrm{STI}}}}\delta_{\mathrm{O},\mathrm{WST}_{\overline{\mathrm{STI}}}}$).

Separate UA surveys carried out across the country allow direct
estimation of HIV prevalences among IDU (source k), FSW (sources k and
q) and non-STI clinic attending migrant subgroups (source l). As
particularly the UA survey covering the $\mathrm{CRB}_{\overline{\mathrm{STI}}}$
population is suspected to suffer from underreporting bias, this is
specifically utilized to inform a lower bound for the corresponding
prevalence parameter.

SOAP records are likely to underestimate HIV prevalence, due to an
opt-out policy in place on HIV testing in STI clinics across the
Netherlands. Information on HIV prevalence in STI clinic users is
limited to those individuals actually submitting to HIV testing, while
only information on attendance is retained from the remaining
patients. Since reluctance to submit to HIV testing is indicative of a
higher risk of HIV infection [\citet{vdbduk08}], it is reasonably
assumed that STI clinic users opting out of HIV testing are more
likely to be HIV positive. Details on how opt-in and opt-out
contributions to HIV prevalence parameters are decoupled and modeled
are given in Section \ref{sssecmix}. UA surveys in Amsterdam (DWAR;
source g) and Rotterdam (ROTan; source v) are also included into the
network of evidence, as they inform HIV prevalence among all STI
clinic users (regardless of ethnicity) in urban areas.

Last, very little information exists on HIV prevalence affecting
low-risk subgroups. Two indirect anonymized sources can be identified:
the national antenatal screening program (source m), which monitors
seroprevalence in pregnant women across the Netherlands in 2007; and
the national registry of blood donors (Sanquin Foundation, source w),
which keeps records of HIV infections among new and regular donors in
the Netherlands in 2007.

Data on blood donors are unlikely to provide unbiased evidence on HIV
prevalence in the low-risk group, as blood donors are at especially
low risk of contracting HIV. Moreover, information from Sanquin is not
categorized by either gender or region, so it captures HIV prevalence
at a very coarse subgroup level. Equally problematic, data from the
national antenatal screening program, which are broadly classified by
ethnicity, provide evidence on HIV prevalence on a population subgroup
not explicitly contemplated by the model, but rather resulting from a
mixture of female subgroups in $\mathcal G$. An assumption of equal
representativeness, in terms of risk group composition, of pregnant
women with respect to the wider female population is introduced to
allow modeling of this indirect (``mixed'') evidence
[\citet{presdean08}].
\subsection{Proportions diagnosed with HIV}
\label{ssecdelta}

Many data sources already informing HIV prevalence also provide
evidence on the extent of disease diagnosis within the target
subgroups. This information, however, is markedly sparse: sample sizes
are small and coverage does not extend to all region-subgroup
combinations, notably excluding MSM, $\mathrm{MGR}_{\overline{\mathrm{STI}}}$ and
$\mathrm{WST}_{\overline{\mathrm{STI}}}$.

Biases also undermine parts of the evidence base. For instance, UA
evidence is known to underestimate $\delta_{r,\mathrm{IDU}}$, due to the
especially hard-to-reach nature of this subgroup. Data are therefore
assumed to inform a lower bound for corresponding
proportions. Similarly, DWAR records on all HIV infections diagnosed in
STI clinics in Amsterdam, due to intrinsic design limitations, provide
a downward-biased estimate of $\delta_{\mathrm{A},\mathrm{STI}}$.
\subsection{Diagnosed HIV infections}
\label{ssecmu}

\setcounter{footnote}{1}

The HIV Monitoring Foundation (SHM; source h) compiles and maintains a
registry of (almost\footnote{In reality, not every diagnosed HIV case
makes timely (if any) contact with national treatment facilities.})
all diagnosed HIV cases in specialized care in the Netherlands,
classified by socio-demographic factors. Absolute counts from the
relevant registry inform the regional risk group composition of
(mixtures of) prevalent HIV diagnoses: namely,
%
\begin{equation}
\mu_{r,g}=N_{r}\rho_{r,g}\pi_{r,g}\delta_{r,g},
\label{eqmi}
\end{equation}
which form a set of functional parameters (see Figure \ref{figinf})
involving all basic parameters of interest. The risk subgroup
classification adopted by SHM does not coincide with that in the MPES
model, since it includes\vadjust{\goodbreak} mixed pregnant women, $\mathrm{SSA}$, $\mathrm{CRB}$, $\mathrm{WST}$ and
unclassified individuals (none in $\mathcal G$) as well as MSM and
IDU. Additionally, cross-matching with records from the Schorer
Monitor reveals an underreporting bias affecting SHM records on
prevalent MSM cases diagnosed across the country. Finally, SHM is also
known to underreporting IDU cases in Amsterdam.

\section{The MPES model structure}
\label{secmod}

The above array of data on HIV prevalence in the Netherlands is
synthesized in a Bayesian statistical model relying upon suitably
chosen standard distributions, in the spirit of case studies already
documented in the literature
[e.g., \citet{adsu06};
\citet{gouad08};
\citet{presdean08}].
\subsection{Sampling distributions}
\label{sseclik}
Count data $x_{r,g}$ from a census- or survey-type study of fixed size
$n_{r,g}$ on subgroup $g\in\mathcal G$ in region $r$ (like, e.g., SOAP
records on HIV diagnoses in STI clinics across the Netherlands) and
characterized by a generic probability parameter $\lambda_{r,g}$ are
naturally modeled via Binomial likelihoods
\[
x_{r,g}\vert\lambda_{r,g}\sim \operatorname{Bin}(n_{r,g}, \lambda_{r,g}).
\]

The total number of diagnosed HIV cases $m_{r\cdot}$ on SHM record as
in care in region $r$ is assumed to follow a Poisson distribution with
regional rate $\mu_{r\cdot}=\sum_g\mu_{r,g}$, with $\mu_{r,g}$ defined
as in \eqref{eqmi}. At the same time, the subgroup sizes
$m_{r,g}$ within each region are assigned a Multinomial distribution
with size parameter $m_{r\cdot}$ and probability vector
$\bolds\xi_r=(\xi_{r,g};g\in\mathcal G)$ with
$\xi_{r,g}=\frac{\mu_{r,g}}{\mu_{r\cdot}}$, so that
\begin{eqnarray*}
m_{r\cdot}\vert\mu_{r\cdot} & \sim& \operatorname{Poi}(\mu_{r\cdot}),\\
m_{r,g}\vert m_{r\cdot},\bolds\xi_r & \sim&
\operatorname{Multin}(m_{r\cdot},\bolds\xi_r).
\end{eqnarray*}

In practice, as explained in Section \ref{seces}, interest does not
always lie in the (often functional) $\lambda$ or $\xi$ parameters,
but rather in the basic parameters they subsume in their
definition. The relationship between basic and functional parameters
is formally determined by the type of mixed, biased or otherwise
indirect evidence available. Examples are illustrated in the following sections.
\subsubsection{Mixed subgroup modeling}
\label{sssecmix}

By classifying individuals into risk\break groups other than those being
modeled, registry-type records provide information on proportions of
diagnosed cases in each risk category (i.e., ratios of the form
$\mu_{r,g}/\sum\mu_{r,g}$, rather than $\delta_{r,g}$), possibly on
suitably defined mixtures of subgroups in $\mathcal G$.

This is, for instance, the case with SHM which, as outlined in Section~%
\ref{ssecmu}, poses a number of modeling challenges. Unclassified
individuals within its records are distributed proportionately across
modeled risk groups, in line with \citet{presdean08}. Additionally,
records on mixed migrant subgroups are modeled by STI clinic
attendance status via the likelihood term
\[
m_{r,g}\vert m_{r\cdot},\xi_{r,g}\sim
\operatorname{Bin}(m_{r\cdot},\xi_{r,g})
\]
for $g\in\{\mathrm{SSA},\mathrm{CRB}\}$, where
\[
\xi_{r,\mathrm{SSA}}  =\frac{\mu_{r,{\mathrm{SSA}}_{\mathrm{STI}}}+\mu_{r,\mathrm{SSA}_{\overline
{\mathrm{STI}}}}}{\mu_{r\cdot}}
\]
and
\[
\xi_{r,\mathrm{CRB}}
=\frac{\mu_{r,\mathrm{CRB}_{\mathrm{STI}}}+\mu_{r,\mathrm{CRB}_{\overline{\mathrm{STI}}}}}{\mu_{r\cdot}},
\]
where $\xi_{r,g}$ denotes the fraction of reported
regional HIV diagnoses in the
$\mathrm{SSA}=\mathrm{SSA}_{\mathrm{STI}} \cup \mathrm{SSA}_{\overline{\mathrm{STI}}}$ and
$\mathrm{CRB}=\mathrm{CRB}_{\mathrm{STI}} \cup \mathrm{CRB}_{\overline{\mathrm{STI}}}$ subgroups.

As explained in Section \ref{ssecpi}, estimation of the HIV
prevalence $\pi^{\mathrm{out}}_{r,g}$ unobserved in subgroups opting out of HIV
testing requires some modeling assumption. Here it is assumed that
prevalence among STI clinic users declining an HIV test would be at
least that of patients with the same risk profile but not submitting
to the test [\citet{vdbduk08}]. This is formalized for
$g\in\{\mathrm{SSA}_{\mathrm{STI}}, \mathrm{CRB}_{\mathrm{STI}}, \mathrm{WST}_{\mathrm{STI}}\}$ through the decomposition
\begin{eqnarray*}
\pi_{r,g} & =&\frac{\#\mathrm{infections}}{N_r\rho_{r,g}}\\
& =&\frac{\#\mathrm{opt\mbox{-}in
\ infections}}{N_r\rho_{r,g}}+\frac{\#\mathrm{opt\mbox{-}out
\ infections}}{N_r\rho_{r,g}}\\
& =&\pi^{\mathrm{in}}_{r,g}+\pi^{\mathrm{out}}_{r,g},
\end{eqnarray*}
where the HIV prevalence $\pi^{\mathrm{in}}_{r,g}$ among those
submitting to the diagnostic test is the parameter actually being
captured by regional SOAP statistics.

In Section \ref{ssecpi} it was also mentioned that, in addition to
SOAP, DWAR and ROTan provide independent information on HIV prevalence
and proportions diagnosed among STI clinic users of
ethnicity in Amsterdam and Rotterdam, respectively. These
aggregate-level data are retained into the model to estimate
corresponding parameters for $r\in\{A,R\}$, that is,
\begin{eqnarray*}
\tilde\pi_{r,\mathrm{STI}} & =&\frac{\sum_{g\in \mathrm{STI}}\rho_{r,g}\pi_{r,g}}{\sum
_{g\in \mathrm{STI}}\rho_{r,g}},\\
\tilde\delta_{r,\mathrm{STI}} & =&\frac{\sum_{g\in
\mathrm{STI}}\rho_{r,g}\pi_{r,g}\delta_{r,g}}{\sum_{g\in
\mathrm{STI}}\rho_{r,g}\pi_{r,g}}.
\end{eqnarray*}

Data on low-risk women from the national antenatal screening program
are dealt with similarly.
\subsubsection{Bias modeling}
\label{sssecbias}

In general, any sample estimating some basic parameter $\vartheta$
with bias $\Delta\neq0$ can be regarded as providing indirect
evidence, on some suitable scale, on $\vartheta$ through the
functional parameter $\psi(\vartheta)=\vartheta+\Delta$. An example of
biased evidence from the case study at hand is offered by CBS
immigration records which, as explained in Section \ref{ssecrho}, do
not include illegal entries and are not classified by STI clinic
attendance status. Letting $\gamma_{r,g}$ indicate the proportion of
\textit{legal} migrants in region $r$ with ethnicity
$g\in\{\mathrm{SSA},\mathrm{CRB}\}$, CBS provides unbiased information on the relative
size $\tilde\rho_{r,g}$ of immigrant subpopulations legally living in
the Netherlands: in functional terms,
\[
\tilde\rho_{r,\mathrm{SSA}}  =\gamma_{r,\mathrm{SSA}}(\rho_{r,\mathrm{SSA}_{\mathrm{STI}}}+\rho
_{r,\mathrm{SSA}_{\overline{\mathrm{STI}}}})
\]
and
\[
\tilde\rho_{r,\mathrm{CRB}}
=\gamma_{r,\mathrm{CRB}}(\rho_{r,\mathrm{CRB}_{\mathrm{STI}}}+\rho_{r,\mathrm{CRB}_{\overline{\mathrm{STI}}}}).
\]
Since no auxiliary data are available to inform the number of illegal
immigrants, either overall or by ethnicity, it is assumed that the
proportions $1-\gamma_{r,g}$ of SSA-born (CRB-born) illegal migrants
in each region ranged between 10\% and 20\%\footnote{Unlike immigrants
from Sub-Saharan African countries, most individuals from the
Caribbean are actually entitled lawful entry into the Netherlands.}
(0\% and 5\%) across the country [\citet{mvvmeet09}].

Furthermore, downward-biased SHM records on MSM and IDU (see Section~\ref{ssecmu}) are modeled by dividing parameters $\mu_{r,\mathrm{MSM}}$ and
$\mu_{\mathrm{A},\mathrm{IDU}}$ with underreporting proportions, in turn separately
estimated from the Schorer Monitor and Amsterdam Cohort Study,
respectively.
\subsection{Prior distributions}
\label{ssecprior1}

Within a Bayesian framework, parameters of a~statistical model are
given some prior probability distribution reflecting the imperfect
knowledge around them. In the present work basic parameters are
typically assigned vague prior distributions. At times, lack of
information around a significant number of parameters
required introducing more structured priors to express either known
constraints or expert opinion.
\subsubsection{Parameter constraints}
\label{ssseccns}

A relatively simple example of the need for a constraining prior
distribution was offered by data from the Sanquin Foundation. As
pointed out in Section \ref{ssecpi}, HIV prevalence among blood
donors is expected to be significantly lower than among the wider
$\mathrm{WST}_{\overline{\mathrm{STI}}}$ subgroup. This is accommodated within the model by
assuming that information from blood donors provides a \textit{lower
bound} $\pi^{\mathrm{L}}_{\mathrm{WST}_{\overline{\mathrm{STI}}}}$ for $\mathrm{WST}_{\overline{\mathrm{STI}}}$ HIV
prevalence at a national
level, where
\[
\pi^{\mathrm{L}}_{\mathrm{WST}_{\overline{\mathrm{STI}}}}\leq\pi_{r,\mathrm{WST}_{\overline{\mathrm{STI}}}}  \qquad \forall r.
\]

These assumptions are introduced to model any sample that
is only known to be biased, but without any additional
information as to the extent of the bias. These data
points are annotated in detail in Table \ref{taba}. In all cases the
modeling structure is naturally
completed by Uniform priors defined over appropriate bounds, as was
done in Section \ref{sssecmix} with the parameters~$\pi^{\mathrm{out}}_{r,g}$. It is then assumed that
$\pi_{r,\mathrm{WST}_{\overline{\mathrm{STI}}}}\leq\min_{g\in\mathcal G}\pi_{r,g}$,
implying that
$\mathrm{WST}_{\overline{\mathrm{STI}}}$ prevalences should not exceed that exhibited by
any other
subgroup in the same region. Last, diagnosed HIV prevalences in any
STI clinic-attending subgroup are conservatively assumed to be at
least 20\%, to prevent unrealistically low parameter estimates.
\subsubsection{Expert opinion}
\label{sssecblf}


Sometimes subjective prior distributions were eli\-cited from
collaborating epidemiologists. This was the case with parameters
characterizing low-risk individuals and, more broadly, outside
Amsterdam and Rotterdam (see Section \ref{ssecpi}). Similarly to
\citet{gouad08} and \citet{presdean08}, let $\pi_{r,g}^s$
denote HIV
prevalence among male and female ($s=\mathrm{m},\mathrm{f}$) individuals with risk
profile $g\neq \mathrm{MSM},\mathrm{FSW}$ in region $r$; the male-to-female prevalence
log-odds ratio is then defined as
\[
\eta_{r,g}=\operatorname{logit}\pi^{\mathrm{m}}_{r,g}-\operatorname{logit}\pi^{\mathrm{f}}_{r,g}.
\]
A two-stage hierarchical model is formulated for prevalence log-odds
ratios: in the first level these are pooled across subgroups to
produce shrunk estimates $\bar\eta_r$; the second then pools regional
log-odds ratios $\bar\eta_r$ across the Netherlands to derive an
overall estimate $\bar{\bar\eta}$. The complete model specification is
thus given by
%
\begin{eqnarray}
\label{eqhrc}
\eta_{r,g}\vert\bar\eta_r,\sigma_r & \sim&\mathcal
N(\bar\eta_r,\sigma^2_r),\nonumber
\\[-8pt]
\\[-8pt]
\bar\eta_r\vert\bar{\bar\eta},\tau,\sigma_r & \sim&\mathcal
N(\bar{\bar\eta},\tau^2).
\nonumber
\end{eqnarray}

Vague hyperpriors on the national log-odds ratio ($\bar{\bar\eta}$)
and on the regional~($\sigma_r$) as well as national ($\tau$) standard
deviations, respectively, measuring the degree of between-subgroup and
between-region heterogeneity among prevalence log-odds ratios,
complete the hierarchical model structure.

While absolute HIV prevalences should not be reasonably expected to be
distributed homogeneously across subgroups within each region,
corresponding male-to-female log-odds ratios can instead be more
plausibly thought of as arising from a common region-specific
distribution, as implied by \eqref{eqhrc}. Shrinkage toward a
regional mean allows information available around some subgroups to
supplement that around others poorly informed; see,
for example, \citet{gelhil07} and annotated bibliography for a
comprehensive
review of the concept of ``borrowing strength.''

Finally, expert opinion helps: to categorize individuals from
HIV-endemic countries by legal entry status when modeling respective
subgroup sizes (as described in Section \ref{sssecbias}); and to
infer HIV prevalences $\pi^{\mathrm{out}}_{r,g}$ among STI clinic-attending
subgroups declining HIV testing (as seen in Section~\ref{sssecmix}). Additional assumptions relating to migrants and STI
clinic users are motivated by the expectation of a higher proportion
of HIV diagnoses among STI clinic users, relative to nonusers with
the same ethnicity and sexual orientation. In more formal terms,
\begin{eqnarray*}
\delta_{r,\mathrm{SSA}_{\mathrm{STI}}} & \geq&\delta_{r,\mathrm{SSA}_{\overline{\mathrm{STI}}}},\\
\delta_{r,\mathrm{CRB}_{\mathrm{STI}}} & \geq&\delta_{r,\mathrm{CRB}_{\overline{\mathrm{STI}}}}
\end{eqnarray*}
and
\[
\delta_{r,\mathrm{MSM}_{\mathrm{STI}}}  \geq\delta_{r,\mathrm{MSM}_{\overline{\mathrm{STI}}}}.
\]
In a similar fashion, the MPES model also includes the
constraints
\[
\delta_{r,\mathrm{CRB}_{\mathrm{STI}}}  \geq\delta_{r,\mathrm{SSA}_{\mathrm{STI}}}
\]
and
\[
\delta_{r,\mathrm{CRB}_{\overline{\mathrm{STI}}}}
\geq\delta_{r,\mathrm{SSA}_{\overline{\mathrm{STI}}}}.
\]
The above are motivated by a better integration in the Netherlands of
Caribbean migrants compared to Sub-Saharan Africans, who tend to be
less familiar with HIV treatment facilities and access routes to
health-care services [\citet{mvvwag05}].
\subsection{Model appraisal}
\label{ssecappr}

Recalling notation introduced in Section \ref{seces}, the
standardized deviance of a particular model is defined as
\[
D(\mathbf
y,\bolds\vartheta)=-2\ln\frac{L(\bolds\vartheta;\mathbf
y)}{L(\hat{\bolds\vartheta};\mathbf y)},
\]
where
$L(\hat{\bolds\vartheta};\mathbf y)$ is the likelihood of the
saturated model (where the number of parameters equals the number of
observations), evaluated at the maximum-likelihood estimate of
$\bolds\vartheta$.

The use of the posterior mean deviance
%
\begin{equation}
\label{eqdev}
\bar D(\mathbf y)=\mathbb
E [D(\mathbf y,\bolds\vartheta)\vert\mathbf y ]
\end{equation}
has been suggested by \citet{spibes02} as a measure of goodness of
fit: under standard regularity conditions, if the model is true,
$\mathbb E [\bar D(\mathbf y) ]\approx n$, so that, in
particular, $\mathbb E [\bar D(\mathbf y) ]\gg n$ would be
suggestive of lack of model fit. As the sampling distribution of $\bar
D(\mathbf y)$ is not well understood [\citet{seadean11}], this idea
is here used informally to identify conflicting information on
specific parameters (see Section \ref{ssecrol}), through the
decomposition $\bar D(\mathbf y)=\sum_{i=1}^n\bar D(\mathbf
y_i)$ of the deviance \eqref{eqdev} into the individual
contributions $\bar D(\mathbf y_i)$ made by each data point
$\mathbf y_i, i=1,\ldots,n$. The fact that, for a true model,
$\mathbb E [\bar D(\mathbf y_i) ]\approx1,$ can be
used to identify specific data points responsible for a potential lack
of fit and to investigate the likely inconsistency in the information
they provide.


\section{Results}
\label{secres}

The MPES model was fitted to the collection of surveillance and survey
data via McMC simulation using the \texttt{WinBUGS} statistical
package [\citet{lun00}]. The code and data required to produce model
estimates are provided as 
 \ref{supbugs}. The
sampling algorithm was started at three independent initial states,
with convergence ascertained by both visual and formal diagnostic
means [\citet{gelrub92}] after 30,000 iterations. After thinning, a
30,000-sized sample from the full posterior distribution was
subsequently retained for drawing inferences.
\subsection{Model inferences}
\label{ssecinf}

Point and interval estimates around basic population and HIV-related
parameters are presented by risk group for the Amsterdam area in Table
\ref{tabinfA}, together with predicted number of HIV infections
classified by diagnosis status; inferences for the remaining
georgraphic areas across the Netherlands are presented as
\ref{supinfronl}.

Most posterior distributions tend to concentrate around parameter
values regarded as plausible by the collaborating epidemiologists,
usually with a reasonable level of accuracy, given the uncertainty
affecting the underlying data. In particular, predicted numbers of
prevalent ($N_r\rho_{r,g}\pi_{r,g}$) and undiagnosed
($N_r\rho_{r,g}\pi_{r,g}(1-\delta_{r,g})$) cases---the key inputs to
health-care decision-making---appear in line with expectations and
concur with results from alternative analytical frameworks
[\citet{vvnprs11}].

Estimates in general reflect the varying accuracy of regional
collection networks, as well as local patterns of subgroup
composition: the precision of inferences can be seen to broadly
decrease when moving from urban concentrations (like Amsterdam, Table
\ref{tabinfA}) into smaller subgroups/areas across the Netherlands
(see 
\ref{supinfronl}). This is, for
instance, the case with estimated proportions of prevalent MSM cases
diagnosed outside Amsterdam and Rotterdam ($\hat\delta_{\mathrm{O},\mathrm{MSM}}$, see
\ref{supinfronl}), which are also significantly lower than those
from the two main cities (see, e.g., $\hat\delta_{\mathrm{A},\mathrm{MSM}}$ from Table~\ref{tabinfA}). Due to the lack of studies targeting
$\delta_{\mathrm{O},\mathrm{MSM}}$, this can only be inferred indirectly from diagnosed
HIV cases ($\mu_{\mathrm{O},\mathrm{MSM}}$) and diagnosed prevalence
($\pi_{\mathrm{O},\mathrm{MSM}}\delta_{\mathrm{O},\mathrm{MSM}}$), the latter in turn being informed by
two biased studies. The uncertainty around resulting MPES estimates is
just a consequence of the synthesis between such scarce evidence and
the mild ranking assumptions on $\delta_{r,g}$ detailed in Section
\ref{sssecblf}.

Estimated proportions of prevalent IDU cases diagnosed in Amsterdam
($\delta_{\mathrm{A},\mathrm{IDU}}$ from Table \ref{tabinfA}) are higher than elsewhere
in the Netherlands ($\delta_{r,\mathrm{IDU}}$ for $r\neq A$ from 
\ref{supinfronl}). Critical appraisal of
$\hat\delta_{\mathrm{A},\mathrm{IDU}}$ is complicated by the large number of data
sources involved. Direct data-based estimates (Table \ref{taba}) can
be seen to be much lower than those produced by the MPES model (Table
\ref{tabinfA}). At the same time, however, the number $\mu_{\mathrm{A},\mathrm{IDU}}$
of diagnosed HIV infections, which suffers from underreporting (see
Section \ref{ssecmu}), disagrees with direct information\vadjust{\goodbreak} separately
available on its building blocks~$\pi_{\mathrm{A},\mathrm{IDU}}$ and $\delta_{\mathrm{A},\mathrm{IDU}}$
(see Section \ref{ssecdelta}). On the other hand, evidence listed in
Table~\ref{taba} on $\pi_{\mathrm{A},\mathrm{IDU}}$ and $\mu_{\mathrm{A},\mathrm{IDU}}$, while biased, is
overall firmer than that around $\delta_{\mathrm{A},\mathrm{IDU}}$ and therefore weighs
more in the synthesis process. In broad terms, records on diagnosed
infections can be seen as an ``anchor'' to the balance between
$\pi_{\mathrm{A},\mathrm{IDU}}$ and $\delta_{\mathrm{A},\mathrm{IDU}}$: by definition, the same number of
diagnosed HIV infections $\mu_{r,g}$ can be obtained with different
combinations of prevalent cases $N_r\rho_{r,g}\pi_{r,g}$ and fractions
diagnosed $\delta_{r,g}$. Since available information allows for more
accurate estimation of $\pi_{\mathrm{A},\mathrm{IDU}}$ compared to $\delta_{\mathrm{A},\mathrm{IDU}}$, the
MPES model reconciles conflicting evidence around $\mu_{\mathrm{A},\mathrm{IDU}}$ by
favoring larger estimates of the more uncertain $\delta_{\mathrm{A},\mathrm{IDU}}$ over
correspondingly lower values of~$\pi_{\mathrm{A},\mathrm{IDU}}$.
\section{Discussion}
\label{secdisc}

Recent applied work has consolidated the role of MPES as a modeling
framework for the estimation of epidemiological indicators of
infectious diseases
[\citet{ad03};
\citet{wltads05};
\citeauthor{adwel06} (\citeyear{adwel06,adwel08});
\citet{gouad08};
\citet{presdean08};
\citet{deanswe09}].
This has paved the way for governmental institutions (e.g., the Medical
Research Council, the National Institute for Health and Clinical
Excellence and the Health Protection Agency in the UK) and
international bodies (e.g., the World Health Organization and UNAIDS)
to increasingly rely on formal evidence synthesis as an analytic tool
to advance epidemiological understanding and support medical
decision-making. The present work represents an additional step toward
expansion of the range of applicability of the MPES approach, as it
illustrates the experience of HIV prevalence estimation in the
Netherlands, a western European country with a concentrated HIV
epidemic and reasonably consolidated and accessible HIV specialist
care.

While relatively extensive in terms of geographic and behavioral
coverage, the array of surveillance and survey data made available by
the Dutch National Institute for Public Health and the Environment
required a comprehensive reappraisal at an evidence synthesis
stage. Problems in the evidence body were identified through an
informal use of deviance statistics in terms of conflicting, biased or
insufficient data on certain region-subgroup
combinations. Inconsistencies thus detected were mostly resolved by
using additional data and/or expert beliefs provided by collaborating
epidemiologists. Nevertheless, some evidence of conflict remained, as
indicated by the overall mean posterior deviance of 258.139, compared
to a total of 186 observations.\footnote{This differs from the nominal
total of 194 (see Section \ref{secdata}) in that it excludes overly
sparse samples---like those from SOAP leading to 0 or 1 maximum
likelihood estimates of some $\delta_{r,g}$ parameters---not
meeting the regularity conditions mentioned in Section
\ref{ssecappr}.} This conflict is mainly around evidence informing
HIV prevalence among migrant women in the rest of the Netherlands, for
which collection of further information was consequently
recommended. Ultimately, model estimates broadly met the expectations
of the pool of epidemiologists involved in the case study.
\subsection{The role within MPES of direct and indirect evidence}
\label{ssecrol}

In general, the process of amalgamating all knowledge available is
\textit{expected} to produce more accurate inferences than those
resulting from a partial or no synthesis. However, as anticipated in
Section \ref{secintro}, the availability of multiple evidence sources
on given parameters can lead to the utilization of discrepant, if not
conflicting, items of information. These discrepancies typically
originate from an incorrect interpretation of what the data represent
(e.g., unrecognized biases), which consequently are inadequately
modeled. If these inconsistencies remain unresolved, MPES inferences
may be less accurate than those obtained from using direct information
alone. This is because MPES estimates arise as a compromise between
estimates separately informed by direct and indirect evidence only,
the more precise of the two weighting more in the balance. The MPES
approach allows resolution of inconsistencies by explicitly modeling
the conflicting items of evidence [e.g., by accounting for biases
in the data, as in \citet{adcli02};
\citet{presdean08}]. In practice, this is
achieved via an interactive reappraisal process, involving the
statisticians and collaborating epidemiologists, of the data sources
flagged by the MPES model as conflicting. Ultimately, any unresolved
conflict of evidence on some parameter would be symptomatic of the
need for additional information---either in the form of field data or
of expert opinion---to be collected.

\begin{table}[b]
\tabcolsep=0pt
\caption{Posterior mean deviances (computed from direct items
of evidence only) and posterior medians with 95\% credibility
intervals for selected basic parameters, obtained from modeling
available direct evidence only, indirect evidence only and all
evidence}
\label{tabdev}
\begin{tabular*}{\textwidth}{@{\extracolsep{\fill}}lcccc@{}}
\hline
  &   & \multicolumn{3}{c@{}}{\textbf{Inferences (\%)}}\\[-5pt]
  &   & \multicolumn{3}{c@{}}{\hrulefill}\\
 \textbf{Parameter}&\textbf{Deviance} & \textbf{Direct} & \textbf{Indirect MPES} & \textbf{Full MPES}\\
\hline
$\pi^{\mathrm{m}}_{\mathrm{R},\mathrm{CRB}_{\mathrm{STI}}}$ & 1.664 & \hphantom{1}1.194 (0.416,
2.598)\tabnoteref{fn:bias}
& 1.393 (1.000, 2.194)\hphantom{0} & 1.492 (1.014, 2.451)\\
$\pi^{\mathrm{f}}_{\mathrm{R},\mathrm{IDU}}$ & 1.887 & 14.489 (8.389, 22.488) & 7.461 (4.198,
13.250) & 10.060 (6.612, 15.030)\\
$\rho_{\mathrm{A},\mathrm{MSM}}$ & 2.102 & \hphantom{1}9.372 (7.456, 11.557) & 13.750 (10.240,
21.240) & 10.750 (9.173, 12.560)\\
$\pi^{\mathrm{f}}_{\mathrm{O},\mathrm{CRB}_{\overline{\mathrm{STI}}}}$ & 8.463 & 1.515 (0.529, 3.292) &
0.191 (0.164, 0.222)\hphantom{0} & 0.194 (0.166, 0.226)\\
$\pi^{\mathrm{m}}_{\mathrm{O},\mathrm{SSA}_{\overline{\mathrm{STI}}}}$ & 9.346 & \hphantom{1}1.116 (0.162, 3.658)\tabnoteref
{fn:bias} & 3.838 (2.228, 6.214)\hphantom{0} & 3.926 (2.227, 6.345)\\
\hline
\end{tabular*}
\tabnotetext[a]{fn:bias}{{Direct evidence is known to be up- or down-ward biased.}}
\end{table}

%
\begin{sidewaystable}
\tabcolsep=0pt
\tablewidth=\textwidth
\caption{Posterior medians with 95\% credibility intervals of
epidemiological parameters and total ($N\rho\pi$) and
undiagnosed ($N\rho\pi(1-\delta)=N\rho\pi-\mu$) infections from
the MPES model of HIV prevalence in Amsterdam}
\label{tabinfA}
{\fontsize{8.6}{10.6}\selectfont{
\begin{tabular*}{\textwidth}{@{\extracolsep{\fill}}lcccccc@{}}
\hline
\textbf{Group} & \textbf{Subgroup} & $\bolds{\hat\rho}$ \textbf{(\%)} & $\bolds{\hat\pi}$ \textbf{(\%)}
& $\bolds{\hat\delta}$ \textbf{(\%)} & $\bolds{\widehat{N\rho\pi}}$ & $\bolds{\widehat{N\rho\pi}-\hat\mu}
$\\\hline
\multirow{3}{*}{{MSM}} & {{STI}} & 0.879 (0.844, 0.913) &
29.100 (24.910, 33.800) & 93.510 (87.810, 97.210) & 726 (617, 846) & 46
(19, 91) \\
& {Non-STI} & 9.871 (8.298, 11.680) & 9.641 (8.074, 11.590) &
88.730 (74.900, 95.370) & 2\mbox{,}682 (2\mbox{,}380, 3\mbox{,}208) & 301 (113, 800) \\
& {All} & 10.750 (9.173, 12.560) & 11.250 (9.645, 13.180) &
89.640 (78.130, 95.570) & 3\mbox{,}404 (3\mbox{,}138, 3\mbox{,}931) & 351 (140, 854) \\
\multirow{3}{*}{{IDU}} & {M} & 0.332 (0.256, 0.399) &
21.365 (15.690, 28.100) & 90.700 (69.660, 99.630) & 198 (132, 286) & 18
(0, 76) \\
& {F} & 0.093 (0.074, 0.108) & 27.260 (21.180, 35.520) & 93.280
(73.751, 99.700) & 71 (54, 95) & 4 (0, 22) \\
& {MF} & 0.212 (0.173, 0.248) & 22.650 (17.840, 28.590) &
90.790 (72.441, 99.610) & 270 (198, 362) & 24 (0, 90) \\[3pt]
{FSW} & {F} & 2.620 (2.561, 2.679) & 3.133
(1.192, 6.252) & 33.975 (4.950, 69.159) & 233 (88, 467) & 148 (43, 367)
\\
\multirow{3}{*}{$\mathrm{WST}_{\mathrm{STI}}$} & {M} & 2.008 (1.958, 2.061) &
0.297 (0.184, 0.577) & 57.630 (26.940, 91.490) & 16 (10, 32) & 7 (0,
23) \\
& {F} & 2.318 (2.265, 2.376) & 0.168 (0.112, 0.317) & 64.910
(32.590, 93.410) & 11 (7, 20) & 3 (0, 12) \\
& {MF} & 2.164 (2.126, 2.202) & 0.234 (0.153, 0.397) & 59.070
(32.140, 90.579) & 28 (18, 48) & 11 (1, 31) \\
\multirow{3}{*}{$\mathrm{SSA}_{\mathrm{STI}}$} & {M} & 0.091 (0.080, 0.102) &
3.732 (2.686, 6.541) & 79.315 (47.120, 97.290) & 9 (7, 16) & 1 (0, 8) \\
& {F} & 0.056 (0.048, 0.065) & 7.348 (5.947, 9.797) & 83.700
(68.730, 97.040) & 11 (10, 15) & 1 (0, 4) \\
& {MF} & 0.073 (0.067, 0.081) & 5.179 (4.160, 7.164) & 80.580
(59.801, 96.700) & 21 (17, 29) & 4 (0, 11) \\
\multirow{3}{*}{$\mathrm{CRB}_{\mathrm{STI}}$} & {M} & 0.315 (0.295, 0.337) &
0.600 (0.452, 1.119) & 84.500 (54.521, 97.860) & 5 (4, 10) & 0 (0, 4) \\
& {F} & 0.271 (0.252, 0.290) & 0.917 (0.767, 1.340) & 87.630
(73.252, 97.890) & 7 (6, 10) & 0 (0, 2) \\
& {MF} & 0.293 (0.279, 0.307) & 0.763 (0.624, 1.108) & 85.360
(66.271, 97.380) & 12 (10, 18) & 1 (0, 5) \\
\multirow{3}{*}{$\mathrm{SSA}_{\overline{\mathrm{STI}}}$} & {M} & 3.825 (3.590,
4.071) & 1.899 (1.186, 3.278) & 64.050 (35.030, 88.019) & 206 (129,
356) & 73 (17, 222) \\
& {F} & 3.394 (3.161, 3.587) & 3.503 (2.877, 4.371) & 72.300
(59.431, 83.320) & 336 (279, 415) & 93 (48, 164) \\
& {MF} & 3.607 (3.430, 3.781) & 2.671 (2.142, 3.531) & 68.670
(51.800, 81.030) & 546 (441, 719) & 170 (88, 342) \\
\multirow{3}{*}{$\mathrm{CRB}_{\overline{\mathrm{STI}}}$} & {M} & 10.970 (10.650,
11.280) & 0.442 (0.292, 0.693) & 75.720 (46.760, 91.440) & 137 (90,
214) & 32 (9, 106) \\
& {F} & 12.910 (12.550, 13.270) & 0.375 (0.294, 0.479) & 80.270
(66.860, 92.130) & 137 (107, 175) & 26 (9, 55) \\
& {MF} & 11.940 (11.680, 12.200) & 0.407 (0.319, 0.539) &
77.410 (59.160, 89.000) & 276 (216, 365) & 61 (26, 144) \\
\multirow{3}{*}{$\mathrm{WST}_{\overline{\mathrm{STI}}}$} & {M} & 71.710 (69.840,
73.360) & 0.065 (0.040, 0.129) & 80.750 (39.700, 98.710) & 132 (81,
264) & 24 (1, 152) \\
& {F} & 78.350 (77.890, 78.800) & 0.066 (0.033, 0.109) & 85.020
(54.701, 98.800) & 147 (72, 242) & 21 (1, 81) \\
& {MF} & 75.030 (74.070, 75.880) & 0.067 (0.042, 0.107) &
82.180 (49.150, 98.530) & 284 (180, 455) & 48 (3, 216) \\
\multirow{3}{*}{{Total}} & {M} & 100 (100, 100) & 1.459
(1.335, 1.679) & 86.655 (75.320, 93.860) & 4\mbox{,}143 (3\mbox{,}791, 4\mbox{,}768) & 553
(234, 1\mbox{,}170)\\
& {F} & 100 (100, 100) & 0.340 (0.286, 0.422) & 67.075 (53.860,
78.700) & 965 (812, 1\mbox{,}200) & 316 (174, 550) \\
& {MF} & 100 (100, 100) & 0.901 (0.831, 1.017) & 82.690
(73.560, 89.200) & 5\mbox{,}120 (4\mbox{,}720, 5\mbox{,}777) & 885 (512, 1\mbox{,}521)\\
\hline
\end{tabular*}
}}
\vspace*{-15pt}
\end{sidewaystable}

As explained in Section \ref{ssecappr}, examination of the
contribution $\bar D(\mathbf y_i)$ provided by each data point
$\mathbf y_i$ to the posterior mean deviance \eqref{eqdev} allows
identification of conflicts between direct and indirect evidence: for
a given item of direct evidence, the farther from 1 its contribution
to \eqref{eqdev}, the more marked the discrepancy of the information
it provides on a particular parameter with the remaining available
evidence. This is illustrated in Table \ref{tabdev}, which for
selected basic parameters reports the posterior mean deviance
contributions to~\eqref{eqdev}, based on their respective direct-only
evidence, alongside corresponding inferences obtained from separately
utilizing direct, indirect and full information. The benefit of full
evidence synthesis in the presence of broadly agreeing sources of
information is made obvious by estimates of~$\pi^{\mathrm{m}}_{\mathrm{R},\mathrm{CRB}_{\mathrm{STI}}}$ and
$\pi^{\mathrm{f}}_{\mathrm{R},\mathrm{IDU}}$: respective MPES inferences combine direct and
indirect evidence, whose consistency is highlighted by deviance
statistics close to~1, to produce narrower credibility intervals than
those arising from a direct approach. Moderately discrepant
information is instead reconciled within the MPES model through a
compromise between direct- and indirect-only inferences: this, as in
the case of $\rho_{\mathrm{A},\mathrm{MSM}}$, yields a credibility interval not
significantly narrower than its direct counterpart, since it conveys
not only the uncertainty within, but also the variability between, the
items of evidence it involves.

The same rationale applies to MPES inferences on parameters informed
by conflicting evidence: similarly to $\rho_{\mathrm{A},\mathrm{MSM}}$, MPES credibility
intervals around~$\pi^{\mathrm{f}}_{\mathrm{O},\mathrm{CRB}_{\overline{\mathrm{STI}}}}$ and
$\pi^{\mathrm{m}}_{\mathrm{O},\mathrm{SSA}_{\overline{\mathrm{STI}}}}$ offer a compromise between the direct
and indirect information separately contributing to their
estimation. The synthesized inferences, however, are clearly less
accurate than their respective direct versions, due to the extent of
the inconsistency undermining the information on
$\pi^{\mathrm{f}}_{\mathrm{O},\mathrm{CRB}_{\overline{\mathrm{STI}}}}$ and $\pi^{\mathrm{m}}_{\mathrm{O},\mathrm{SSA}_{\overline{\mathrm{STI}}}}$,
as also indicated by the correspondingly large deviance statistics. In
this case, while seemingly offering no immediate advantage over a
direct method, the deviant MPES estimates point to those parts of the
evidence body which remain inconsistent, hence suggesting what type of
supplementary information would be most useful to resolve the
discrepancy.

Furthermore, it can be seen that the leverage each evidence source
applies to the final MPES estimate is determined by its corresponding
sample size, not by it being graded as ``direct'' or ``indirect'': this is
shown in Table \ref{tabdev} with~$\pi^{\mathrm{f}}_{\mathrm{O},\mathrm{CRB}_{\overline{\mathrm{STI}}}}$,
whose estimate is more driven by the stronger indirect
evidence. Finally, it is worth noting that the availability, in a MPES
setup, of different sets of inferences, each resulting from a
different level of evidence synthesis, further stresses the robustness
and efficiency of a full MPES approach over any arbitrary selection of
items from the complete evidence base. Eventually the MPES approach
contributed to a better understanding of the nature of those evidence
conflicts which, pending the availability of additional data (the
provision of which will be discussed for future updates of national
estimates), remain unresolved.
\subsection{Prior information in MPES}
\label{ssecprior2}

As detailed in Section \ref{ssecprior1}, the presented MPES model
relies on a number of prior assumptions. This is in line with the MPES
spirit of informing the analysis with \textit{all} available evidence,
not just ``hard'' data. Often reliance on expert opinion is regarded as
inappropriate, in that if misused it could steer the analysis toward
partly subjective outcomes. On the other hand, as notably pointed out
in \citeauthor{rob07} [(\citeyear{rob07}), Chapter~1], knowledge does
not exclusively derive from
field data, but actually builds on it. Substantive prior  information
informing the illustrated case study was typically introduced
pragmatically: earlier versions of the MPES model featuring
fewer/milder prior assumptions than those listed in Section
\ref{ssecprior1} produced overly inaccurate (i.e., with unduly wide
credibility bounds) estimates for some poorly informed subgroup-region
combinations.\footnote{Results not shown.} An MPES model can help
in identifying those parameters whose estimation would benefit the most
from the collection of larger/additional samples. To this end, while
MPES modeling falls short of indicating which design strategy would
yield largest efficiency gains, insights in this respect are naturally
offered by more formal decision-theoretic tools, such as those based
on the concept of value of information
[\citet{parino09}, Chapter 13]. While these are receiving increasing
attention by the environmental and health sciences community, they
remain the subject of ongoing investigation and fall outside the remit
of this paper.
\subsection{Current HIV prevalence estimation platforms}
\label{ssecUNAIDS}

The MPES approach lends itself as a valuable framework for national
HIV prevalence estimation. Alternative options have been freely made
available in recent years by the UNAIDS Reference Group on Estimates,
Modeling and Projections: that is, the Estimation and Projection
Package [EPP, \citet{epp04};
\citet{epp06}] and the Workbook Method
[\citet{wb04};
\citet{wb05}], each implemented in a bespoken software package
(unlike MPES).


EPP assumes the national population is subdivided into nonoverlapping
risk subgroups, for which historical records of size and HIV
prevalence are available. EPP then fits a simple transmission model to
the prevalence data via Sampling Importance Resampling from a Bayesian
Melding perspective [\citet{pooraf00}], generating a cluster of
epidemic curves for each urban/rural and subgroup-specific
sub-epidemic [\citet{alkraf07};
\citet{rafbao10}]. Resulting national HIV
prevalence and incidence projections can then be fed into the
stand-alone Spectrum module [\citet{spc04}] to predict over time the
number of individuals living with HIV or AIDS, new HIV infections,
etc.

The Workbook Method estimates and projects HIV prevalence in countries
lacking an HIV surveillance network consistently monitoring local
prevalence patterns over time. Similarly to MPES and EPP, albeit to a
coarser degree, Workbook estimates rely on a classification of the
target population by risk profiles for which values of the maximum and
minimum size and of HIV prevalence are available. The various
combinations of lower-upper bounds are then cross-multiplied and
averaged to obtain informal ``plausibility'' ranges for national HIV
prevalence. This in turn can be imported into EPP/Spectrum to obtain a
wider array of ancillary HIV epidemic descriptors.

A comparative discussion of the advantages and shortfalls of the three
approaches (MPES, EPP/Spectrum, Workbook) is presented elsewhere
[\citet{vvnprs11}]. Extensive criticism of Workbook estimates has led
to a~marked shift toward utilization and development of EPP among
epidemiologists and practitioners in the field. While MPES has been
only recently extended to HIV prevalence estimation, its flexibility
shows promise for application to increasingly varied and complex data
structures. Successful implementations have been carried out to
incorporate time-series data for the estimation of HIV prevalence and
incidence trends and patterns among MSM in England and Wales
[\citet{prsdea11}]. Additional case studies to be conducted via MPES
modeling are currently being sought among eastern European countries,
since this should facilitate the continuing development necessary for
the methodology to reach higher levels of dissemination and maturity.

\section*{Acknowledgments}
The authors gratefully acknowledge all institutions, data managers and
researchers who supplied data as well as valuable discussion and
follow-up. The authors are also grateful to three anonymous referees
for their helpful advice and comments.


%
\begin{supplement}[id=supdataro]
\sname{Supplement A}
\stitle{HIV prevalence data in Rotterdam and the rest of the
Netherlands}
\slink[doi]{10.1214/11-AOAS488SUPPA}
\slink[url]{http://lib.stat.cmu.edu/aoas/488/Supplement_A.ps}
\sdatatype{.ps}
\sdescription{Surveillance- and sur\-vey-type data supporting HIV
prevalence estimation Rotterdam and the Rest of the Netherlands.}
\end{supplement}
%
%
\begin{supplement}[id=supbugs]
\sname{Supplement B}%
\stitle{MPES model and data files\\}%
\slink[doi,text={10.1214/11-AOAS488SUPPB}]{10.1214/11-AOAS488SUPPB}%
\slink[url]{http://lib.stat.cmu.edu/aoas/488/Supplement\%20B.zip}%
\sdatatype{.zip}%
\sdescription{\texttt{WinBUGS} code of the MPES model and data
inputs enabling HIV prevalence estimation in the Netherlands.}
\end{supplement}
%
%
\begin{supplement}[id=supinfronl]
\sname{Supplement C}%
\stitle{HIV prevalence estimates in Rotterdam and the Netherlands
(including and excluding Amsterdam)\\}%
\slink[doi,text={10.1214/11-AOAS488SUPPC}]{10.1214/11-AOAS488SUPPC}%
\slink[url]{http://lib.stat.cmu.edu/aoas/488/Supplement\%20C.ps}%
\sdatatype{.ps}%
\sdescription{Posterior inferences on HIV prevalence descriptors by
risk subgroup in Rotterdam and the Netherlands (separately
including and excluding Amsterdam and Rotterdam).}%
\end{supplement}

%

\printaddresses

\end{document}